\documentclass[useAMS,usenatbib]{mn2e}

\newcommand{\vN}{\mbox{N}}

\newcommand{\vnh}{\hat{\mbox{\bf {n}}}}

\usepackage{multirow}
         
\def\plotancho#1{\includegraphics[width=18cm]{#1}}
\usepackage{graphicx} 
\usepackage{epsfig}
\usepackage{amsmath} 
\usepackage{txfonts}
\def\refe@jnl#1{{#1}}
\def\aj{\refe@jnl{Astron.~J.}}                  
\def\araa{\refe@jnl{Annu.~Rev.~Astron.~Astrophys.}}
\def\apj{\refe@jnl{Astrophys.~J.}}                 
\def\apjl{\refe@jnl{Astrophys.~J.~Lett.}}          
\def\apjs{\refe@jnl{Astrophys.~J.~S.~S.}}          
\def\aap{\refe@jnl{Astron.~Astrophys.}}            
\def\mnras{\refe@jnl{Mon.~Not.~R.~Astron.~Soc.}}   
\def\prd{\refe@jnl{Phys.~Rev.~D}}        
\def\fcp{\refe@jnl{Fund.~Cos.~Phys.}}  
\def\physrep{\refe@jnl{Phys.~Rep.}}
\def\physlett{\refe@jnl{Phys.~Lett.}}

\DeclareTextSymbol{\degre}{OT1}{23}

\title[ISW--tSZ: ISW extraction out of CMB data only]{The ISW-tSZ cross correlation: ISW extraction out of pure CMB data}
\author[N. Taburet et al.]{N. Taburet\thanks{E-mail:
    nicolas.taburet@ias.u-psud.fr}$^1$,
    C. Hern\'andez-Monteagudo$^2$, N. Aghanim$^1$, M. Douspis$^1$, R.A. Sunyaev$^{2,3}$
  \\ $^1$ Institut d'Astrophysique Spatiale, Universit\'e Paris-Sud 11 \& CNRS (UMR 8617), B\^at.
  121, 91405 Orsay Cedex, France
  \\$^2$ Max-Planck Institut f\"ur Astrophysik,
  Karl-Schwarzschild-Str. 1, 85740 Garching, Germany
  \\$^3$ Space Research Institute, Russian Academy of Sciences, Profsoyuznaya 84/32, 117997 Moscow, Russia}
\begin{document}

\date{}

\pagerange{\pageref{firstpage}--\pageref{lastpage}} \pubyear{2009}

\maketitle
\label{firstpage}

\begin{abstract}

 If Dark Energy introduces an acceleration in the universal expansion then
large scale gravitational potential wells should be shrinking, causing a
blueshift in the CMB photons that cross such structures (Integrated
Sachs-Wolfe effect, [ISW]). Galaxy clusters are known to probe those
potential wells. In these objects, CMB photons also experience inverse
Compton scattering off the hot  electrons of the intra-cluster medium, and
this results in a distortion with a characteristic spectral signature of
the CMB spectrum (the so-called thermal Sunyaev-Zel'dovich effect, [tSZ]).
Since both the ISW and the tSZ effects take place in the same potential
wells, they must be spatially correlated. We present how this cross
ISW-tSZ signal can be detected in a {CMB-data} contained way {by} using the frequency
dependence of the tSZ effect in multi frequency CMB experiments like
{\it Planck}, {\em without} requiring the {use} of external large scale structure tracers data.
{We find that by masking low redshift clusters, the shot noise level decreases significantly, 
boosting the signal to noise ratio of the ISW--tSZ cross correlation. We also find that galactic and
extragalactic dust residuals must be kept at or below the level of $\sim 0.04$ ($\mu$K)$^2$ at $\ell=10$,
a limit that is a factor of a few below {\it Planck}'s expectations for foreground subtraction. If this is achieved,
CMB observations of the ISW-tSZ cross correlation should also provide an independent probe
for the existence of Dark Energy and the amplitude of density perturbations.}

\end{abstract}

\begin{keywords}
cosmology: theory -- methods: statistical -- cosmic microwave
background  -- galaxies: clusters: general
\end{keywords}

\section{Introduction}

The primary Cosmic Microwave Background (CMB) and especially its angular
power spectrum provides us with powerful constraints on the content of
the universe and its evolution.  It is now well established that an
accurate understanding of the primary CMB power spectrum requires a
good understanding of the secondary CMB anisotropies resulting from
the interaction of the CMB photons with the matter along the line of
sight from the last scattering surface to the observer \citep[see][for a review]{Aghanimrevue08}. The great
efforts {undertaken} to understand these secondary anisotropies, in order
to best recover the primary CMB, also provide us with powerful
independent cosmological probes when the secondary anisotropies are
regarded as a source of information rather than contamination.

Among those secondary CMB anisotropies, some result from the
gravitational interaction of the CMB photons with the potential wells
they cross. {One of them is the} Integrated Sachs-Wolfe (ISW)
effect, {by which CMB photons experience some blue/redshift as}
they pass through large scale time evolving potential wells
\citep{SachsWolfe1967}. Since a dark energy like component is expected
to affect the growth of large scale structures, {making them
  shallower}, {a} detection of the ISW effect is an important
probe {for} establishing {its} existence - {provided that} the
Universe is flat and general relativity is a correct description of
gravity - and constraining the equation of state of such a component.

 Detection claims of the ISW effect arose as soon as the first
  year WMAP data were released. Those
  claims were based upon cross correlation analyses of WMAP CMB data
  and {galaxy density templates built from different surveys}. While
  most of the first analyses were conducted in real space (i.e., by
  computing the angular cross-correlation function), subsequently new
  results based upon Fourier/multipole and wavelet space were
  presented. The results from the WMAP team on the cross correlation
  of NRAO Very Large Sky Survey (NVSS) with WMAP data
  \citep{Nolta2004} were soon followed by other analyses applied not
  only on NVSS data, but also on X-ray and optical based catalogs like
  HEA0, SDSS, APM or 2MASS,
  \citep{Boughn2003-4,Fosalba2003,Scranton2003,Fosalba2004,Afshordi2MASS}. As
  subsequent data releases from both the CMB and the SDSS side became
  public, new studies prompted further evidence for significant cross
  correlation between CMB and Large Scale Structure (LSS) data, e.g.,
  \citet{padmanabhan2005,Cabre2006,Giannantonio2006,Rassat2007}. By
  that time, wavelet techniques were also applied on NVSS and WMAP
  data, providing the highest significance ISW detection claims at the
  level of 3--4 $\sigma$,
  \citep{Vielva2006,Pietrobon2006,McEwen2007}. The initial effort of
  \citet{Granett2008}, consisting in stacking voids and superclusters
  extracted from SDSS data, yielded a very high significance ($\sim
  4\sigma$) ISW detection claim. However, it was later found in
  \citet{Granett2009} that such signal could not be due to ISW only,
  since a gravitational potential reconstruction from the Luminous Red
  Galaxy (LRG) sample of SDSS yielded a much lower signal ($\sim
  2.5\sigma$). \citet{Giannantonio2008} and \citet{Ho2008} used
  different LSS surveys in a combined cross-correlation analysis with
  CMB data and claimed high significance ($\sim 4-5 \sigma$) ISW
  detections.  

However, doubts on the validity of such claims have also arised
recently.  \citet{chm2006} first pointed out the lack of significant
cross correlation between WMAP 1st year data and density surveys built
upon 2MASS, SDSS and NVSS surveys on the large angular scales, but
detected the presence of radio point source emission and thermal
Sunyaev-Zel'dovich effect on the small scales. In \citet{chm2008} a
study of the expected signal-to-noise ratio (S/N) for different sky
coverages was presented, and it was found that in the standard
$\Lambda$CDM scenario the ISW -- density cross correlation should be
well contained in the largest angular scales, ($l<50-60$). This was
proposed as a consistency check for ISW detection against point-source
contamination. In \citet{chm2009} cross-correlation analyses between
NVSS and WMAP 5th year data provided no evidence for cross-correlation
in the large angular range ($l<60$). A signal at the 2--3 $\sigma$
level was however found at smaller scales, although its significance
increased with increasing flux thresholds applied on NVSS sources (in
contradiction with expectations for the ISW probed by NVSS and raising
the issue of radio point source contamination). Furthermore, the
intrinsic clustering of NVSS sources on the large scales (relevant for
the ISW) was found too high for the commonly assumed redshift
distribution for NVSS sources, as found in previous works,
\citep{Negrello2006,Raccanelli2008}. Regarding ISW detection claims
based upon SDSS data, there is also some ongoing discussion after
recent failures in finding any statistical significance for the ISW,
\citep{Bielby2008,Martin2010,Tom2010}. This situation is partially
caused by the fact that the ISW is generated on the large angular
scales and at moderate to high redshifts ($z\in [0.1,1.3]$). Deep
galaxy surveys covering large fractions of the sky are hence required
to sample the ISW properly, but those are not available yet (or not
properly understood).\\

  Ideally one would try to find the ISW contribution to CMB
  anisotropies by using CMB data {\em exclusively}. And in this
  context the thermal Sunyaev-Zel'dovich effect becomes of
  relevance.  The thermal Sunyaev-Zel'dovich (tSZ) effect
\citep{SZ72} results from the inverse Compton scattering of the CMB
photons off the galaxy clusters electrons, and is expected to provide
cosmological constraints on the normalisation at 8$h^{-1}$ Mpc of the
density fluctuations power spectrum, $\sigma_8$, as well as on the
amount of matter $\Omega_m$ and to a lower extent on the
dark energy equation of state \citep[e.g.][]{BattyeWeller03}. Since
both the tSZ and the ISW effects probe large scale structures and
their evolution, a correlation is therefore expected between these two
signals. {As \citet{chmSunyaev2005} pointed out, provided that the
  tSZ has a definite and well known frequency dependence, it is
  possible to combine different CMB maps obtained at different
  frequencies in search for a {\em frequency dependent} ISW--tSZ
  cross-correlation. The advantages of this approach are two folded:
  {\it (i)} only CMB data (obtained at different frequency channels)
  are required, and hence there is no need for using and characterising
  an external LSS catalog, and {\it (ii)} a better handle on
  systematics is provided since the ISW--tSZ cross correlation has a
  perfectly known frequency dependence that can be searched for in
  multiple channel combinations. This approach in experiments like
  {\it Planck} \footnote{{\it Planck} URL site:\\ {\tt
      http://www.sciops.esa.int/index.php?project=PLANCK}}, covering
the whole sky in a wide frequency range, is well suited to separate the
tSZ from other components present in the microwave range.

After introducing in Section 2 the theoretical background, we 
  study in Section 3 a combination of CMB maps at different
frequencies that provides an unbiased estimate of the ISW-tSZ
angular power spectrum. We also compute the expected
  significance for the ISW--tSZ cross correlation.
In Section 4 we analyze
how the large scale structure contributes to the tSZ and ISW auto
power spectra in different redshift ranges.
 This allows us to define a strategy to optimize the S/N of
  the ISW--tSZ cross-correlation by applying a selective mask on galaxy
  clusters. We then investigate in Section 5 the limitations of our
  method due to the presence of galactic and extragalactic foregrounds and
  suggest some approaches to minimize their impact. We present our
  conclusions in Section 6.

\section{Angular power spectra of large scale structures tracers}
CMB photons can interact with the matter situated along the line of sight from the last scattering surface to the observer, and thus suffer gravitational or scattering effects. This produces fluctuations of the observed CMB temperature on the sky. The most important scattering effect is known as the Sunyaev-Zel'dovich effect that arises when CMB photons scattered off the electrons of the intra cluster gas. Since these electrons are lying within the large scale structure potential wells, a correlation is thus expected between the SZ effect and the ISW temperature fluctuations due to the energy change of the CMB photons that pass through time evolving potential wells.

\subsection{Late ISW}
{In a $\Lambda$CDM scenario, the accelerated expansion of the universe makes large scale potential wells to shrink. As a consequence, CMB photons crossing a potential well $\phi$ do not loose as much energy exiting this well as what they gained when falling into it}. This is known as the late ISW effect and results in a modification of the CMB blackbody temperature \citep[e.g.,][]{enrique}:
\begin{equation}
\frac{\Delta T}{T_{\rm CMB}}=-\frac{2}{c^2}\int \mathrm{d}\eta\frac{\mathrm{d}\phi}{\mathrm{d}\eta},
\end{equation}
where $\eta$ is the conformal time {defined as $d\eta = dt/a(t)$, where $t$ and $a(t)$ are the coordinate time and the scale factor in a standard Friedmann-Robertson-Walker (FRW) metric.}
The relation between the gravitational potential and matter distribution variations is given by the Poisson equation in physical coordinates :
\begin{equation}
\nabla^2\phi=4\pi G\bar\rho_{\rm m}(1+\delta).
\end{equation}
This equation is easier to solve in the comoving frame and in Fourier space ($k$ stands for the wavenumber and the subscript c denotes comoving units):
\begin{equation}
\label{eq:phi_k}
\phi(k)=-\frac{3}{2}H_0^2\Omega_{\rm m0}\frac{\delta_{k_{\rm c}}}{ak_{\rm c}^2},
\end{equation}
where $H_0$ is the Hubble constant. We have used the relation between the critical density of the Universe and the expansion rate of the homogeneous background $\rho_{\rm c}(a)=(3H^2(a))/(8\pi G)=\rho_{\rm m}(a)/\left(\Omega_{\rm m0}a^{-3}\left(\frac{H_0}{H(a)}\right)^2\right)$.

In the linear regime, as long as the different $k$ modes are not coupled to each other, the matter overdensity for a pressureless
fluid can be written $\delta_k(a)=D_+(a)\delta_k(z=0)$ where the growing mode $D_+(a)$ is a solution to the following differential equation :
\begin{equation}
\label{eq:structgrowth}
\ddot\delta(k)+2H\dot\delta(k)=4\pi G\bar\rho_{\rm m}\delta(k),
\end{equation}
in which dots denote a derivation with respect to the physical time.

Using Eq. (\ref{eq:phi_k}) we can express the CMB temperature fluctuations due to the ISW in the linear regime :
\begin{equation}
\frac{\Delta T}{T_{\rm CMB}}=\frac{3H_0^2}{c^2}\Omega_{\rm m0}\int{\frac{a\mathrm{d}r_{\rm c}}{c}\frac{\mathrm{d}}{\mathrm{d}t}\left(\frac{D_+(a)}{a}\right)\int{\frac{\mathrm{d}^3k_{\rm c}}{(2\pi)^3}\frac{\delta_{k,z=0}}{k_{\rm c}^2}e^{-i{\bf{k_{\rm c}}}.{\bf{r_{\rm c}}}}}}.
\end{equation}
The temperature fluctuations due to the ISW effect can be projected on the spherical harmonics basis,
\begin{equation}
\label{eq:almISW}
a_{\ell m}^{\rm ISW}=4\pi(-i)^\ell\int{\frac{\mathrm{d}^3k_{\rm c}}{(2\pi)^3}Y_{\ell m}^*({\bf{k_{\rm c}}})\Delta_\ell^{\rm ISW}(k_{\rm c})\delta_{k_{\rm c},z=0}},
\end{equation}
with the ISW transfer function :
\begin{equation}
\label{eq:transferFuncmISW}
\Delta_\ell^{\rm ISW}(k_{\rm c})=\frac{3H_0^2}{c^3}\Omega_{\rm m0}\int{\mathrm{d}r_{\rm c}a\frac{\mathrm{d}}{\mathrm{d}t}\left(\frac{D_+(a)}{a}\right)\frac{1}{k_{\rm c}^2}j_\ell(k_{\rm c}r_{\rm c})}.
\end{equation}

\subsection{The thermal SZ effect}
The SZ effect consists of two terms. The main one is the thermal SZ effect (tSZ) that is due to the inverse Compton scattering of the CMB photons off the intracluster gas hot electrons. The second one is the kinetic SZ effect (kSZ)  which is a Doppler shift due to galaxy clusters' motion with respect to the CMB rest frame. The tSZ effect transfers some energy from the hot electrons to the CMB photons. As a result, in the direction of the cluster, the CMB intensity is decreased in the Rayleigh Jeans part of the spectrum and increased in the Wien part.
This translates into a characterisitic spectral signature $g_\nu$ of the induced CMB temperature secondary fluctuations, expressed as a function of the adimensional frequency $x=h\nu/(k_{\rm B}T_{\rm e})$ and the comptonisation parameter $y$. Neglecting relativistic corrections, this parameter can be written as 
\begin{equation}
\frac{\Delta T}{T_{\rm CMB}}=g_\nu y=\left(x\frac{e^x+1}{e^x-1}-4\right)y.
\end{equation}
The comptonization parameter $y$ corresponds to the integrated electronic pressure along a given line of sight through the cluster :
\begin{equation}
y=\frac{k_{\rm B}\sigma_{\rm T}}{m_{\rm e}c^2}\int dl n_{\rm e}T_{\rm e}.
\end{equation}
The kSZ, on the other hand, does not have a different spectral signature from the CMB. The secondary temperature fluctuations power spectrum due to the kSZ is about 2 to 4 times smaller
than the one induced by the tSZ at 150 GHz \citep[][]{Sehgal2010,LuekerSPT2010}. We can safely neglect its contribution in our analysis as explained in Section 3.

In order to calculate the temperature fluctuations due to a population of N clusters, one can use the halo approach \citep{ColeKaiser88,CooraySheth2002}. In this article, we adopt the line of sight approach that was introduced in \cite{chm2006LOS}.
\begin{equation}
\frac{\Delta T}{T_{\rm CMB}}=g_\nu\int \mathrm{d}r_{\rm p}\sum_i^Ny_{\rm 3D,c}(r_{\rm p})u({\bf{r_{\rm p}}}-{\bf{w_i}})
\end{equation}
where $r_{\rm p}$ stands for physical distances, $w$ for the line of sight position, and $y_{\rm 3D,c}$ is the value of $\frac{k_{\rm B}T_{\rm e}}{m_{\rm e}c^2}\sigma_{\rm T}n_{\rm e}$ at the center of the cluster and $u$ is the electronic radial pressure profile, $y_{\rm 3D}(w)=y_{\rm 3D,c}u(w)$. Replacing the discrete summation with an integral over the position we get :
\begin{eqnarray}
\frac{\Delta T}{T_{\rm CMB}} & = & g_\nu\int \mathrm{d}r_{\rm p}\int\mathrm{d}M y_{\rm 3D,c}(M,r_{\rm p})\int\mathrm{d}w_{\rm p}\left.\frac{\mathrm{d}n}{\mathrm{d}M}\right|_{\rm p}(w_{\rm p},M,r_{\rm p})\nonumber\\
 & & u({\bf{r_{\rm p}}}-{\bf{w_{\rm p}}},M,r_{\rm p})
\end{eqnarray}
The convolution between the mass function and the cluster profile is easier to handle in Fourier space :
\begin{eqnarray}
\frac{\Delta T}{T_{\rm CMB}} & = & g_\nu\int\mathrm{d}r_{\rm p}\int\mathrm{d}M\int\frac{\mathrm{d}^3k_{\rm p}}{(2\pi)^3}\left.\frac{\mathrm{d}n}{\mathrm{d}M}\right|_{\rm p}(k_{\rm p},M,r_{\rm p})\nonumber\\
 & & \tilde y_{\rm 3D}(k_{\rm p},M,r_{\rm p})e^{-i{\bf{k_{\rm p}}}.{\bf{r_{\rm p}}}}
\end{eqnarray}
{Since galaxy clusters are not exclusively Poisson distributed, in the Fourier description of the spatial distribution of these sources a correlation term adds to the Poisson term}. It represents the modulation of the cluster number density by the underlying density field :
\begin{equation}
\left.\frac{\mathrm{d}n}{\mathrm{d}M}\right|_{\rm p}({\bf{k_{\rm p}}},M,r_{\rm p})=\left.\frac{\mathrm{d}\bar n}{\mathrm{d}M}\right|_{\rm p}(M,r_{\rm p})\left[\delta_{\rm D} ({\bf{k_{\rm p}}})+b(M,r_{\rm p})\delta_k\right]
\end{equation}
where the linear bias is modelled with $b(M,z)=1-1/\delta_C+\delta_c/\sigma^2(R,z)$ \citep{MoWhite96}.

{That is, the power spectrum of the tSZ angular anisotropy can be
  written as the contribution of two different terms, \citep[e.g.,][]{KomatsuKitayama99}. The first one refers to the Poisson/discrete nature of these sources, and is known as the 1-halo term. The second one refers to the spatial modulation of the density of these sources, obeying the large scale density field, and is referred to as the 2-halo term, which is sensitive to the underlying matter density distribution.} In this article we are mainly interested in the detection of the cross ISW-tSZ term, to which only this second term contributes.

The temperature fluctuations due to the SZ effect 2halo term can be projected on the spherical harmonics basis,
\begin{equation}
a_{\ell m}^{\rm tSZ2h}=4\pi(-i)^\ell\int{\frac{\mathrm{d}^3k_{\rm p}}{(2\pi)^3}Y_{\ell m}^*({\bf{k_{\rm p}}})\Delta_\ell^{\rm tSZ2h}(k_{\rm p})\delta_{k_{\rm p},z=0}}
\end{equation}
with the tSZ 2-halo term transfer function :
\begin{eqnarray}
\Delta_\ell^{\rm tSZ2h}(k_{\rm p}) & = & g_\nu\int{\mathrm{d}r_{\rm p}\int{\mathrm{d}M\left.\frac{\mathrm{d}\bar n}{\mathrm{d}M}\right|_{\rm p}(M,r_{\rm p})b(M,r_{\rm p})\tilde y_{\rm 3D}(k_{\rm p},M,r_{\rm p})}}\nonumber\\
 & & D_+(r_{\rm p})j_\ell(k_{\rm p}r_{\rm p})
\end{eqnarray}
The Fourier transform of $y_{\rm 3D}(r)$ is :
\begin{equation}
\label{eq:ytilde}
\tilde y_{\rm 3D}(k_{\rm p},M,z)=4\pi\int y_{\rm 3D}(r_{\rm p},M,z)j_0(k_{\rm p}r_{\rm p}) r_{\rm p}^2\mathrm{d}r_{\rm p}
\end{equation}
Switching to comoving units, we write the $a_{\ell m}$ in a form similar to Eqs. (\ref{eq:almISW}) and (\ref{eq:transferFuncmISW}) :
\begin{equation}
\label{eq:almSZ}
a_{\ell m}^{\rm tSZ2h}=4\pi(-i)^\ell\int{\frac{\mathrm{d}^3k_{\rm c}}{(2\pi)^3}Y_{\ell m}^*({\bf{k_{\rm c}}})\Delta_\ell^{\rm tSZ2h}(k_{\rm c})\delta_{k_{\rm c},z=0}}
\end{equation}
with the tSZ 2-halo term transfer function :
\begin{eqnarray}
\label{eq:transferFuncmSZ}
\Delta_\ell^{\rm tSZ2h}(k_{\rm c}) & = & g_\nu\int{a\mathrm{d}r_{\rm c}\int{\mathrm{d}Ma^{-3}\left.\frac{\mathrm{d}\bar n}{\mathrm{d}M}\right|_{\rm c}(M,r_{\rm c})b(M,r_{\rm c})}}\nonumber\\
 & & \tilde y_{\rm 3D}(k_{\rm p},M,r_{\rm c})D_+(r_{\rm c})j_\ell(k_{\rm c}r_{\rm c})
\end{eqnarray}
In the following we have used the \cite{ShethTormen01} mass function, the \cite{KomatsuSeljak2002} model for the intracluster electronic distribution and the WMAP-5 year cosmological parameters \citep{KomatsuWMAP09} : $\Omega_\Lambda=0.721$, $\Omega_{\rm m}=0.279$, $\Omega_{\rm b}=0.046$, $h=0.701$, $\sigma_8=0.817$ and $n_{\rm s}=0.96$.

\subsection{ISW-tSZ Cross power spectrum}
Using Eqs. (\ref{eq:almISW}), (\ref{eq:transferFuncmISW}), (\ref{eq:almSZ}) and (\ref{eq:transferFuncmSZ}) it is straightforward to obtain the cross correlation between the ISW effect and the tSZ effect :
\begin{eqnarray}
\label{eq:ISWSZexact}
C_\ell^{\rm ISW-tSZ} & = & \left<a_{\ell m}^{\rm ISW}{a_{\ell m}^{tSZ2h}}^*\right>\nonumber\\
		& = & \frac{2}{\pi}\int{k_{\rm c}^2\mathrm{d}k_{\rm c}\Delta_\ell^{\rm ISW}(k_{\rm c})\Delta_\ell^{\rm tSZ2h}(k_{\rm c})P(k_{\rm c},z=0)}
\end{eqnarray}
where we have introduced the matter power spectrum :
\begin{equation}
\left<\delta_{{\bf{k}},z=0}\delta^*_{{\bf{k'}},z=0}\right>=(2\pi)^3\delta_{\rm D}({\bf{k}}-{\bf{k'}})P(k,z=0)
\end{equation}
We speed up the computation of the power spectra by adopting the Limber approximation :
\begin{eqnarray}
\label{eq:ISWSZLimber}
C_\ell^{\rm ISW-tSZ} & \simeq & g_\nu\frac{3H_0^2}{c^3}\Omega_{\rm m0}\int{\mathrm{d}z\frac{1}{4\pi}\frac{\mathrm{d}V_{\rm c}}{\mathrm{d}z}P_{\rm m}\left(\frac{\ell+1/2}{\rm D_c},z=0\right)}\nonumber\\
 & & \left(\frac{1}{\ell+1/2}\right)^2a\frac{\partial}{\partial t}\left(\frac{\rm D_+(a)}{a}\right){\rm D_+(a)}\int{\mathrm{d}M\left.\frac{\mathrm{d}\bar n}{\mathrm{d}M}\right|_{\rm c}(M,z)}\nonumber\\
 & & b(M,z)\tilde y_{\rm 3D}\left(\frac{\ell+1/2}{\rm D_A},M,z\right)
\end{eqnarray}
{This Limber approximation is} based on the closing relation of the spherical Bessel function,
\begin{equation}
\int k^2\mathrm{d}kj_\ell(kr_1)j_\ell(kr_2)\mathcal{F}(k)\simeq\frac{\pi}{2}\frac{\delta_{\rm D}(r_2-r_1)}{r_1^2}\mathcal{F}\left(\frac{\ell+1/2}{r_1}\right),
\end{equation}
where $\mathcal{F}$ is a function that slowly varies with $k$. We used $k=\frac{\ell+1/2}{r}$ in order to ensure an error in $\mathcal{O}(\ell^{-2})$ while the frequently employed flat sky approximation $k=\frac{\ell}{r}$ only ensures an error in $\mathcal{O}(\ell^{-1})$ \citep{AfshordiLimber}.

In Fig. \ref{fig:ISWSZ_NoLimber_Limber} we present the CMB-CMB, ISW-ISW, tSZ-tSZ and ISW-tSZ power spectra calculated at 100 GHz without using the Limber approximation. We also represent the instrumental noise for a {\it Planck} like experiment as well as the cosmic variance associated to the CMB.

Fig. \ref{fig:ecrelat_NoLimber_Limber} represents the relative difference between the various angular power spectra calculated using the Limber or the exact formula.
While the 'standard' Limber approximation ($k\simeq\ell/r$) ensures a 10\% precision from $\ell\geq6$ and $\ell\geq20$ for the tSZ-tSZ and ISW-ISW power spectra respectively, these limit multipoles become $\ell\geq3$ and $\ell\geq1$ when using the $k\simeq(\ell+1/2)/r$ approximation. As discussed in \cite{Afshordi2MASS} the errors induced by this later approximation are small compared to the cosmic variance. Therefore in the following we use this $k\simeq(\ell+1/2)/r$ version of the Limber approximation.

\begin{figure}
\includegraphics[width=8cm]{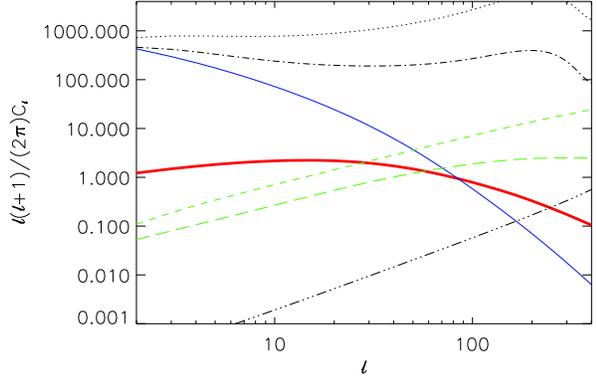}
  \caption{ISW-tSZ angular cross power spectrum at 100 GHz (thick solid red line), ISW angular power spectrum (thin solid blue), tSZ-1halo term (short dashed green), tSZ-2halos term (long dashed green).
The CMB power spectrum is represented as a black dotted line for comparison. The dot-dashed line represents the cosmic variance associated to the CMB and the triple dot-dashed line is the {\it Planck} noise at 100 GHz.}
\label{fig:ISWSZ_NoLimber_Limber}
\end{figure}

\begin{figure}
\includegraphics[width=8cm]{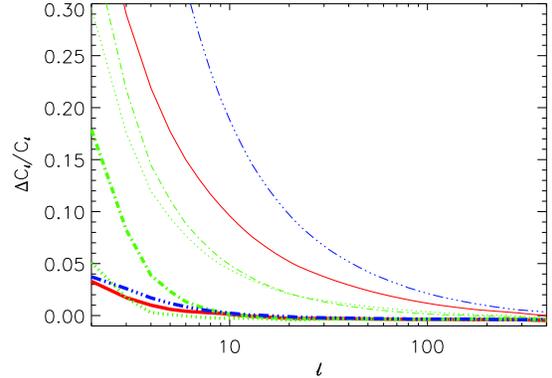}
  \caption{Relative error when using the $k\simeq(\ell+1/2)/r$ Limber approximation (thick lines) or the $k\simeq\ell/r$ Limber approximation (thin lines). The solid red line stands for the ISW-tSZ angular cross power spectrum, the triple dot-dashed blue line for the ISW angular power spectrum, the dot-dashed green line for the tSZ-1halo term and the dotted green line for the tSZ-2halos term. This plot clearly illustrates the great advantage of using the $k\simeq(\ell+1/2)/r$ approximation instead of the 'standard' $k\simeq\ell/r$ Limber approximation in low $\ell$ studies of the power spectra.}
\label{fig:ecrelat_NoLimber_Limber}
\end{figure}

The relative amplitude of the ISW-tSZ power spectrum with respect to the contaminant spectra (primary CMB and tSZ) makes it less difficult to be detected at large scales than at smaller scales (more details in Section \ref{sec:SNRidealcase}, see also \cite{Cooray02ISWLSScorrel,chmSunyaev2005}). For multipoles lower than 100, we see in Fig. \ref{fig:ISWSZ_NoLimber_Limber} that the effect of the instrumental noise can be neglected since its power spectrum is more than one order of magnitude lower than the astrophysical signals considered here.
It is nevertheless essential to notice that the ISW-tSZ signal, we are interested in measuring, is much smaller than the cosmic variance associated to the CMB. In the next section, we present a method, based on \cite{chmSunyaev2005}, that uses the caracteristic spectral signature of the ISW-tSZ signal, to separate it from the CMB.

\section{Detection level of the ISW-SZ cross correlation - ideal case}
\label{sec:SNRidealcase}
\cite{chmSunyaev2005} have shown that using combinations of multifrequency observations of the microwave sky can allow to avoid the limit due to the cosmic variance and unveil a weak signal whose frequency signature differs from the CMB blackbody law.
Following this idea, we propose a combination of different channels in order to unveil the ISW-tSZ signal, in spite of it is dominated by the primary CMB in each channel.
Since the tSZ vanishes at 217 GHz (in the non-relativistic assumption), the channel combination $217*\nu_i-217*217$ ($\nu_i \neq 217$ GHz) should give an unbiased estimate of the ISW-tSZ$_{\nu_i}$ cross power spectrum in the limit in which we neglect the contaminants such as point sources or galactic dust.

In a first analysis we calculate the signal to noise ratio (S/N) of the ISW-tSZ in the ideal case in which the noise is only constituted of primary CMB and tSZ autocorrelations.
The variance of the $C_\ell^{\rm ISW-tSZ_{\nu_i}}$ is :
\begin{equation}
\label{eq:ClISWSZvariance_gal}
\Delta^2(C_\ell^{\rm ISW-tSZ_{\nu_i}})=\langle \left(a_{\ell m}^{217}a_{\ell m}^{\nu_i}-a_{\ell m}^{217}a_{\ell m}^{217}\right)^2\rangle-{\langle a_{\ell m}^{217}a_{\ell m}^{\nu_i}-a_{\ell m}^{217}a_{\ell m}^{217}\rangle}^2
\end{equation}
which reduces, in the ideal case, to
\begin{equation}
\label{eq:ClISWSZvarianceIdeal}
\Delta^2(C_\ell^{\rm ISW-tSZ_{\nu_i}})=\left[C_\ell^{\rm ISW-tSZ}\right]^2+\left(C_\ell^{\rm ISW}+C_\ell^{\rm CMB}+N_\ell\right)\left(C_\ell^{\rm tSZ2h}+C_\ell^{\rm tSZ1h}\right).
\end{equation}
{The kinetic SZ effect has the same spectral dependence as the primary CMB therefore it will be subtracted by the channel combination we propose. Furthermore, the kinetic SZ contribution, at the large scale of interest in this study, is orders of magnitude smaller than the one coming from the primary CMB, it thus does not affect our estimation of the noise term (Eq. \ref{eq:ClISWSZvarianceIdeal}).

For a full sky coverage, the number of independent modes is $2\ell+1$ and can be approximated as $f_{\rm sky}(2\ell+1)$ for a partial sky coverage $f_{\rm sky}$.
We therefore write the signal to noise ratio for the ISW-tSZ cross correlation at multipole $\ell$ as
\begin{equation}
\left.\frac{S}{N}\right|_\ell=\left[\frac{(2\ell+1)f_{\rm sky}\left[C_\ell^{\rm ISW-tSZ}\right]^2}{\left[C_\ell^{\rm ISW-tSZ}\right]^2+\left(C_\ell^{\rm ISW}+C_\ell^{\rm CMB}+N_\ell\right)\left(C_\ell^{\rm tSZ2h}+C_\ell^{\rm tSZ1h}\right)}\right]^{1/2}.
\end{equation}
The cumulative S/N up to multipole $\ell$ for a full sky survey is
\begin{equation}
\frac{S}{N}(\ell)=\left[\sum_{\ell'=2}^\ell\frac{(2\ell'+1)\left[C_{\ell'}^{\rm ISW-tSZ}\right]^2}{\left[C_{\ell'}^{\rm ISW-tSZ}\right]^2+\left(C_{\ell'}^{\rm ISW}+C_{\ell'}^{\rm CMB}+N_{\ell'}\right)\left(C_{\ell'}^{\rm tSZ2h}+C_{\ell'}^{\rm tSZ1h}\right)}\right]^{1/2}.
\end{equation}
In case of partial sky coverage or sky cuts of the Galaxy, the low multipoles are not independent anymore, one should adopt a conservative approach starting the summation from $\ell_{\rm min}\simeq\pi/\theta_{\rm sky}$, $\theta_{\rm sky}$ being the survey size in its smallest dimension \citep{Cooray02ISWLSScorrel}. {Note that in the expression above the $g_{\nu}^2$ factors describing the frequency dependence of the tSZ effect cancel out. {This means that in the ideal case the choice of the frequency $\nu_i$ does not affect the signal to noise level of the ISW-tSZ cross-correlation measurement. Nevertheless, as discussed in section 5, the level of contaminants will depend on the frequency choice which will impact on the signal to noise ratio.}}

In Fig. \ref{fig:SNRideal} we represent the signal to noise ratio contribution from each multipole when considering a cosmic variance limited full sky experiment. The solid and dotted lines show the results obtained when using Eq. (\ref{eq:ISWSZexact}) and the Limber approximation of Eq. (\ref{eq:ISWSZLimber}) respectively. Most of the contribution to the signal to noise ratio comes from multipoles ranging from 5 to 30 with a maximum contribution at $\ell=13$. This multipole range corresponds to the scales in which the ISW-tSZ cross term is expected to reach its maximum amplitude and more importantly dominates over the tSZ autocorrelation term.

\begin{figure}
\includegraphics[width=8cm]{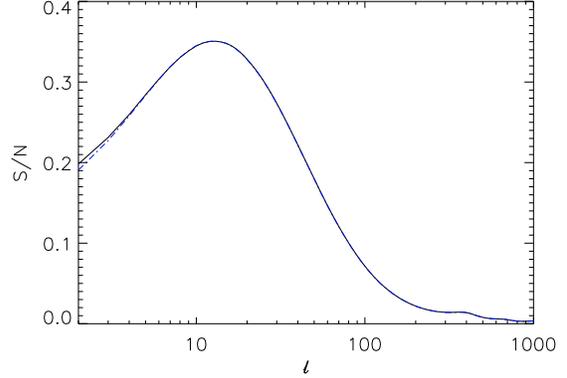}
  \caption{S/N contribution from each multipole for a cosmic variance limited full sky experiment with no contaminants. The solid black is obtained without Limber approximation while the dot dashed blue line is obtained using the Limber approximation as in Eq. (\ref{eq:ISWSZLimber})}
\label{fig:SNRideal}
\end{figure}

The results for both the no Limber and Limber approximations are presented in Fig. \ref{fig:CumulSNRideal}. We found that the cumulative S/N is of the order of 2.2, which is about half the signal to noise ratio found by \cite{Cooray02ISWLSScorrel}.
The main difference between the computation of the angular power spectra in his article and our computation is that we introduced the \cite{KomatsuSeljak2002} cluster pressure profile (Eq. \ref{eq:ytilde}) when estimating $C_\ell^{\rm ISW-tSZ}$ and $C_\ell^{\rm tSZ-tSZ}$.
It is not clear from Eqs. (32) and (33) in \cite{Cooray02ISWLSScorrel} if they account for the cluster pressure profile and what model they use.

\begin{figure}
\includegraphics[width=8cm]{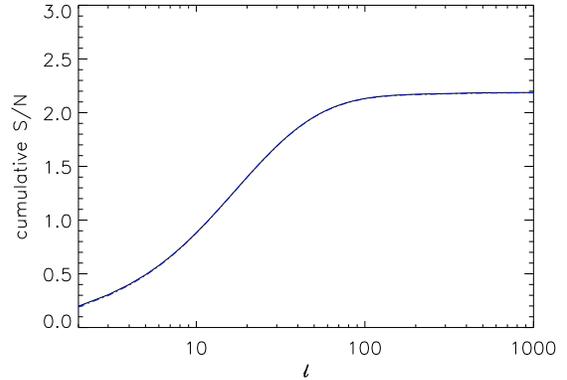}
  \caption{Cumulative signal to noise ratio. same as Fig. \ref{fig:SNRideal}}
\label{fig:CumulSNRideal}
\end{figure}

As pointed out by recent CMB experiments \citep{LuekerSPT2010,KomatsuWMAP7,DunkleyACT2010}, the measured tSZ power spectrum seems to have a lower amplitude than expected from analytical calculations or simulations. Apart from point sources residual that could fill the tSZ decrement observed at frequencies below 217 GHz, this discrepancy could be explained if the $\sigma_8$ value used to compute the cluster abundance in the spectrum modelisation is overestimated. Nevertheless in this case there would be tension between a lower $\sigma_8$ value derived from SZ observations and values derived from primary CMB measurements \citep{KomatsuWMAP7} or X-ray cluster counts \citep{Vikhlinin2009}.
The disagreement between the observed and predicted tSZ power spectrum could also be due to an overestimation of the tSZ effect induced by each individual cluster. 
Theoretical models describing the electronic density and temperature profiles (such as KS02), output from hydrodynamical numerical simulations (such as those obtained by \cite{Sehgal2010}) as well as pressure profiles derived from X-ray observations (such as the one in \cite{ArnaudProfile09}) all seem to predict a higher tSZ {power spectrum} than what is effectively observed. Physical processes such as supernovae or AGN energy feedback to the intra-cluster medium or a non thermal pressure contribution could explain the lower level of observed SZ signal \citep[e.g.][]{Shaw2010}.
Here we used the theoretical KS02 profile to calculate the power spectra (Eq. \ref{eq:ytilde}).
Comparing the effect of various cluster electronic profiles on the amplitude or shape of the ISW-tSZ paper is beyond the scope of this paper and will be investigated in a subsequent work.

The main contribution to the $C_\ell^{\rm ISWtSZ}$ variance in Eq. (\ref{eq:ClISWSZvarianceIdeal}) is the $\left(C_\ell^{\rm ISW}+C_\ell^{\rm CMB}\right)\left(C_\ell^{\rm tSZ2h}+C_\ell^{\rm tSZ1h}\right)$ term, and especially the tSZ contributions that multiply the CMB noise term. We then propose to mask some SZ clusters in order to significantly reduce the noise term without decreasing too much the ISW-tSZ signal.

\section{Optimising the ISW-tSZ detection}

In order to optimise the detection of the ISW-tSZ signal, we need to determine which clusters contribute more to the ISW-tSZ signal and which clusters are responsible for the tSZ-tSZ {shot} noise {on} large scales. We therefore compute the halo contributions to the $C_\ell$s, as a function of their redshift, for different $\ell$ values. We present the results in Fig. \ref{fig:dCldz}. On the large scales (several degrees) we are interested in, the tSZ signal mainly comes from clusters at a redshift lower than 0.3 for the 2-halo term (dot-dashed red line) and 0.03 for the one halo term (dashed blue line). On those same angular scales, the ISW signal (dotted green line) mainly arises from time varying potential wells at redshift [0.2;1]. The cross ISW-tSZ signal mainly comes from clusters at a redshift $z\in [0.06;0.8]$ (solid black line). We therefore expect that masking the low redshift clusters will enhance the signal to noise ratio of the ISW-tSZ.

\begin{figure*}
\includegraphics[width=16cm]{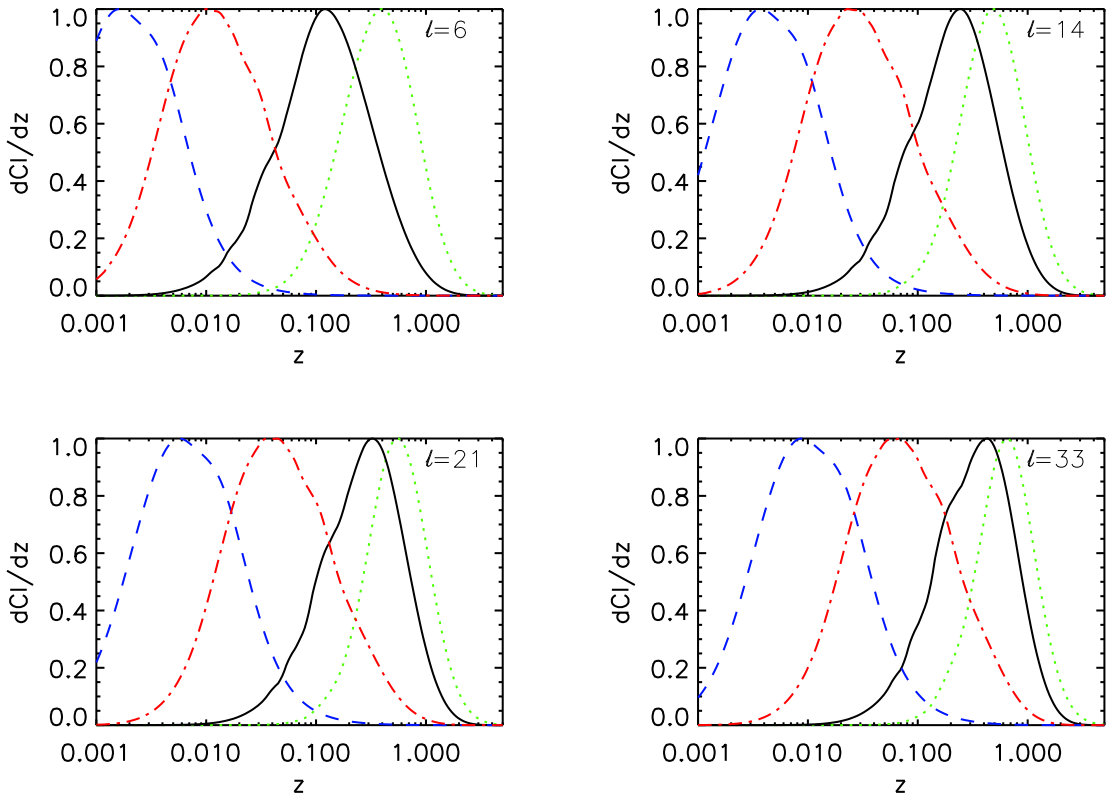}
  \caption{Normalized contribution to the $C_\ell$s as a function of the halos redshift, for different multipoles. The color coding is as follows : dotted green : ISW-ISW, solid black : ISW-tSZ, dot-dashed red : tSZ-tSZ (2halos term), dashed blue : tSZ-tSZ (1halo term).}
\label{fig:dCldz}
\end{figure*}

\subsection{Sharp redshift cut}
We first apply a sharp cut in redshift in order to simulate the effect of masking low $z$ clusters. The power spectra for a $z=0.3$ threshold are presented in Fig. \ref{fig:spectracutz03}. As expected, the 1halo term of the tSZ contribution is strongly decreased, as {is} the 2-halo term to a lower extent (solid green lines). For multipoles lower than 100, the dominant tSZ contribution is now the 2-halo term. Since the ISW-tSZ signal comes from higher redshift than the tSZ signal, masking has a lower impact on its power spectra. The ISW-ISW signal is only reduced at very large scales ($\ell<10$) and this relative variation does not exceed 25\%.

\begin{figure}
\includegraphics[width=8cm]{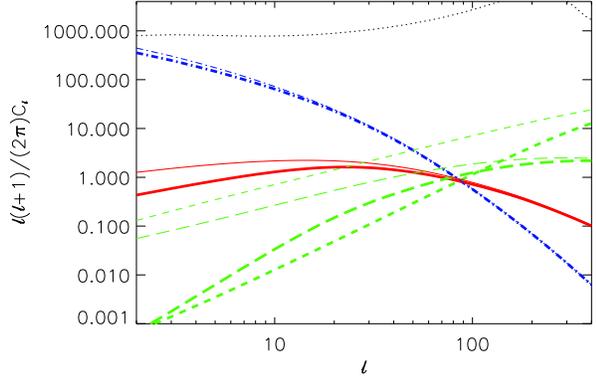}
  \caption{The thick lines represent the power spectra after masking clusters at $z<0.3$ (The color coding is as follows : solid red : ISW-tSZ, dot-dashed blue : ISW-ISW, short dashed green : tSZ-tSZ (1halo term), long dashed green : tSZ-tSZ (2halo term)). For comparison the thin lines represent the contribution due to all clusters. The dotted line is the primary CMB angular power spectrum.
}
\label{fig:spectracutz03}
\end{figure}

As shown in Fig. \ref{fig:correlCoeff_ISWtSZ_cutz03}, the drastic reduction of the tSZ 1-halo term strongly boosts the ISW-tSZ correlation coefficient, which is defined as 
\begin{equation}
r_\ell^{\rm ISWtSZ}=\frac{C_\ell^{\rm ISWtSZ}}{\sqrt{C_\ell^{\rm ISW}C_\ell^{\rm tSZ1h+2h}}}.
\label{eq:corcoef1}
\end{equation}
 This can be easily understood since the tSZ 1-halo, as opposed to the 2-halo term, term does not trace the large scale correlations and behaves as noise for the ISW-tSZ detection. 

\begin{figure}
\includegraphics[width=8cm]{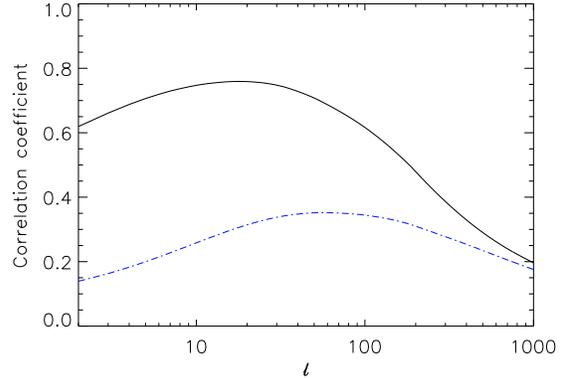}
  \caption{ISWtSZ correlation coefficient. The solid black line is the coefficient after masking $z<0.3$ clusters, which drastically reduces the tSZ 1 halo term at multipoles lower than 100. The dot-dashed blue line is the correlation coefficient without any mask.}
\label{fig:correlCoeff_ISWtSZ_cutz03}
\end{figure}

We then determine what would be the signal to noise ratio that could be obtained for different redshift cut thresholds. Fig. \ref{fig:CumulSNRcut} shows the maximum signal to noise ratio for the ISW-tSZ cross-correlation is reached when masking $z<0.3$ clusters. A simple cut to mask low $z$ clusters allows to increase the S/N from 2.2 to 5.1. Should one also mask $M>5\times 10^{14}$M$_\odot$ clusters only, the significance level would rise to $\sim6\sigma$. {These detection levels compare well to the results of 
\citet{Afshordi04ISWUserManual}, who claims that it should be expected a detection at the level of
 $\sim7.5\sigma$ for an ideal ISW-galaxy correlation. According to this work, this S/N is expected to be of the order of 5 for an all-sky survey with 10 millions galaxies within $0<z<1$, provided the redshift systematic errors are lower than 0.05 and the systematic anisotropies of the survey do not exceed 0.1\%.}
\cite{DouspisISWoptimise} showed that an Euclid like mission should ideally provide a detection of the ISW at a significance level of $5\sigma$. {Note that }
we obtain similar signal to noise ratio with our {approach} but without the need to use non CMB data.

\begin{figure}
\includegraphics[width=8cm]{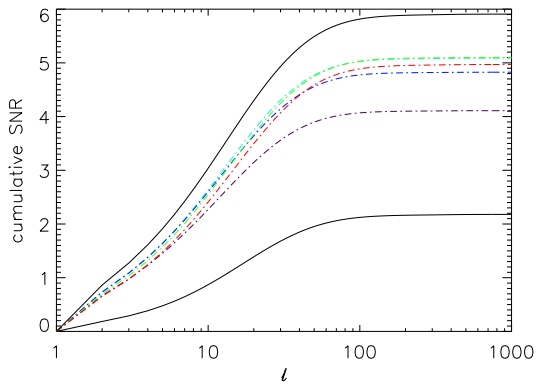}
  \caption{Cumulative signal to noise ratio of the ISW-tSZ detection in the ideal case ($f_{\rm sky}=1$, no foregrounds). The dot-dashed curves represent the S/N obtained for different cut in redshift : 0.1 (purple), 0.2 (blue), 0.3 (light blue), 0.4 (green) and 0.5 (red). As a comparison the solid black curve reminds the S/N obtained without masking any cluster.
The highest solid black curve is the cumulative S/N obtained for a $z<0.3$ and $M>5\times 10^{14}$M$_\odot$ cut.}
\label{fig:CumulSNRcut}
\end{figure}

\subsection{A more realistic approach : using a SZ selection function}
{The previously introduced sharp cuts in redshift assumed that redshifts and masses were available for a relevant subset of clusters. We now relax this assumption } and build a theoretical selection function in order to determine which clusters can be detected -and thus masked- in a {\it Planck}-like survey.

A galaxy cluster is assumed to be detected through its SZ signal if its beam convolved Compton parameter exceeds the confusion noise and its integrated signal $Y$ is higher than $\lambda$ times the instrumental sensitivity simultaneously in the 100, 143 and 353 GHz channels, with $\lambda$ the detection significance level, \citep{Bartelmann01,Taburet09}
The selection functions for 1, 3 and 5 $\sigma$ detections are represented in Fig. \ref{fig:SelecFunc}.

\begin{figure}
\includegraphics[width=8cm]{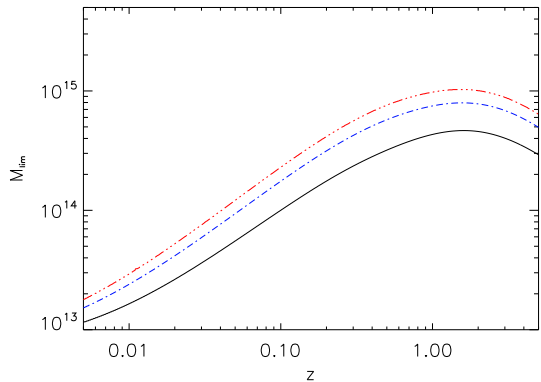}
  \caption{Selection functions for 1 (black solid line), 3 (dot-dashed blue line) and 5 (triple dot dashed red line) $\sigma$ detection of SZ clusters for a {\it Planck}-like survey.}
\label{fig:SelecFunc}
\end{figure}

As shown in Fig. \ref{fig:CumulSNRSelecFunc}, masking the clusters detected with a high detection significance (5 $\sigma$, red line) increases the cumulative signal to noise ratio to 3.7. When masking clusters that are detected at a 3$\sigma$ level, the cumulative S/N is enhanced to 4. 
Reaching this level of detection requires to mask faint clusters and thus suppose a good understanding of the cluster selection function.

\begin{figure}
\includegraphics[width=8cm]{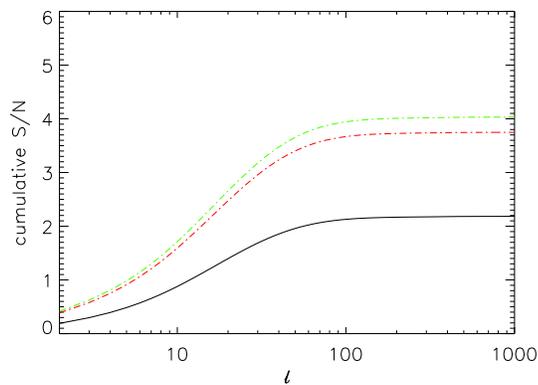}
  \caption{Cumulative signal to noise ratio of the ISW-tSZ detection after masking the 5 (dot-dashed red line) and 3 (dot-dashed green line) $\sigma$ detected SZ clusters. black line reminds the result without masking any cluster.}
\label{fig:CumulSNRSelecFunc}
\end{figure}

\section{Impact of point sources ang galactic residuals}

In the previous section we considered an idealized scenario in which
the only signals were the primary CMB, the ISW, the tSZ and the
instrumental noise. Nevertheless it is of common knowledge that radio
and infrared galaxies as well as our Galaxy are important contributors
to the observed signal at submillimeter wavelengths. For that, we use
state-of-the-art models describing the emission of the
Milky Way a millimeter wavelenghts, together with the contribution from
radio and submillimeter sources. The frequencies for which we built our
foreground model are 100, 143, 217 and 353 GHz, which correspond
 to the HFI instrument of the {\it Planck} mission\footnote{URL site of 
 HFI: {\tt http://www.planck.fr/heading1.html}}. We considered five different contaminants, namely
 free-free, synchrotron and dust emission (coming from our galaxy), and radio 
 and infrared extragalactic sources. For the free-free and synchrotron, we scaled the
 maps produced by the WMAP team at the V and W bands of this experiment 
 (corresponding to 74 and 94 GHz, respectively) to higher frequencies. 
 We used the maps available at the LAMBDA repository\footnote{LAMBDA 
  URL site: {\tt http://lambda.gsfc.nasa.gov}}, and computed the effective spectral
  index in thermodynamic temperature for each pixel between the V and
  the W bands. These spectral indexes were used to extrapolate the thermodynamic 
  temperature to the higher frequencies under consideration. The dust emission 
  induced by the Milky Way was predicted in our frequency range my means of
  model 8 of  \citet[][ hereafter denoted by SFD]{fds}\footnote{The data and
  the software to build this galactic dust map were downloaded from\\
  {\tt http://www.astro.princeton.edu/$\sim$schlegel/dust/cmb/cmb.html} }. This model
  predicts the galactic dust emission down to angular scales of $\sim$ 5 arcmins, which
  suffices in the context of ISW studies. Finally, we used the point source
  maps produced by  \citet{Sehgal2010},
  who modeled the contribution of the extragalactic source population, both in the radio and the
  submillimeter. In that work, a halo catalog resulting from a large cosmological hydrodynamical
  simulation was populated with radio and dusty galaxies, in such a way that the spatial 
  distribution of those sources follows the clustering of dark matter halos. Both
  galaxy populations were designed to meet different constraints obtained from
  radio, infrared and millimeter observations, (see \citet{Sehgal2010} for details). 
  They were projected on high resolution sky maps at different observing frequencies, 
  which differ from HFI channels' central frequencies for a few percent in most of the cases. 

  Our proposed approach to unveil the ISW -- tSZ correlation 
  compares CMB observations obtained at frequencies 
  for which the tSZ is non zero with CMB observations observed at 217 GHz. This
  frequency lies in the Wien region of the CMB black body spectrum, in which the
  CMB brightness drops very rapidly for increasing frequencies and dust contribution
  (both galactic and extragalactic) becomes dominant.  Therefore in a relatively narrow
  frequency range the CMB is surpassed by dust emission, and it is in this frequency range
  where an extrapolation of dust properties (observed at high frequencies) down to lower
   frequencies (where CMB is dominant) must be carried out. On the large angular scales
   of relevance for the ISW it is dust in the Milky Way the main source of contamination,
   and its accurate subtraction is actual critical for our purposes. An experiment like
   HFI counts with frequency channels centered at 353, 545 and 857 GHz, which probe
   the regime where dust emission is well above the CMB contribution. We shall use
   those channels to correct for dust (both galactic and extragalactic) at lower frequencies.
   Our approach attempts by no means to be exhaustive nor systematic, but simply tries to display the degree of accuracy required at subtracting dust emission  in order to unveil the tSZ -- ISW cross correlation in a {\it Planck}-like experiment .
  
  We first built a mask that covered those regions where the Milky Way emission, {\em both}
  in radio and sub-millimeter was stronger. We sorted in intensity (from bigger to 
  smaller values)  the templates of free-free and synchrotron in the V band (as produced by the 
  WMAP team) and the SFD dust  template at 353 GHz. 
  Masking a given level of emission (for instance, the 25\% of brightest pixels) in each
 template yielded two masks that were very similar (particularly in the galactic plane), with differences
 corresponding mostly to high latitude clouds being bright either in the radio or submillimeter 
 (but not on both). The final mask was the product of the two masks built upon the radio and dust templates. The fraction of un-covered sky, $f_{sky}$, was then set as a free parameter in the mask construction. 
 
According to the SFD dust templates, one has to take into account the spatial variation of the effective spectral index if one is to accurately correct for dust emission in the 100 - 217 GHz frequency range. In these templates, an effective spectral coefficient in thermodynamic temperature (defined as the ratio of thermodynamic temperatures between two different channels, i.e., $\alpha_{353,\;j} \equiv \delta T_j / \delta T_{353\;GHz} $) is correlated with the thermodynamic temperature at 353 GHz, as the left panel of Fig. \ref{fig:galaxy1} shows. The curvature at low temperatures is a consequence of the grey body law describing the dust emission in IR galaxies, and we make use of it when subtracting the dust emission at low frequencies. In this low temperature regime, the effective spectral coefficients from IR galaxies differ from the that of the Milky Way, and for this reason a more accurate scaling could be obtained by treating the local cirrus component separately from the extra-galactic IR one. This can be achieved, in high lattitude regions, using H1 data that traces the galactic cirrus emission, in order to remove their contribution.
We sorted pixels {of HEALPix\footnote{HEALPix's URL site: {\tt http://www.healpix.org}} resolution parameter $N_{side}=64$} outside the mask according to their intensity in the dust template at 353 GHz, and binned them in groups of length $n_{groups}$, in each of which a different estimate of $\alpha_{353,j}$ is estimated in the low frequency channels\footnote{ {Unless otherwise specified, this is the pixelization resolution used in our analyses }}.  At these frequencies the observed signal in pixel $\vnh$ is modeled as 
\begin{equation}
T^j(\vnh ) =  N^j(\vnh ) + \alpha_{353,\;j}(\vnh ) M^{353} (\vnh ) + R^j(\vnh),
\label{eq:model_temp}
\end{equation}
where $M^{353}(\vnh ) $ is the dust template at 353 GHz (including both galactic and extragalactic emission), $R^j(\vnh ) $ is the extrapolation from the radio template of WMAP's V band,  and $N^j(\vnh ) $ is what we regard as the noise component at frequency $j$. Most component separation algorithms attempt to fit for all relevant components (including CMB, radio and dust contributions) at each frequency. Now we concern only about the impact of dust residuals, which at these frequencies are dominant. {We find however that by using a mask acting on the 10\% brightest pixels in the synchrotron plus free-free template, the effect of radio remains always below that due to dust}.  
The effective noise $N^j$ contains the residuals of a first order CMB subtraction, and will be assumed to show no spatial correlations (i.e., to be white noise).
For a high level of effective noise it will be necessary to bin the signal in larger groups (bigger $n_{groups}$) in order to measure an average $\alpha_{353,j}$ throughout the bin. However, this average estimate of the effective spectral coefficient in the bin will produce an estimate of the dust contribution at frequency $j$ and pixel $\vnh$ that will {\em not correspond exactly} to the exact value, and this mismatch will be more relevant the wider the bins are (i.e., the bigger $n_{groups}$ is). On the other hand, if noise were negligible one could make $n_{groups}=1$ and measure the dust contribution very accurately at each sky position $\vnh$. 

The residuals from the dust subtraction can be then written as
\begin{equation}
\delta T^j_{res}(\vnh ) =  T^j(\vnh ) - \tilde{\alpha}_{353,\;j} M^{353} (\vnh ),
\label{eq:model_res}
\end{equation}
where the $\tilde{\alpha}_{353,\;j} $ estimate is computed after minimizing 
\begin{equation}
\chi^2 = \sum_{i,l} (T^j-\tilde{\alpha}_{353,\;j} M^{353})_i \vN_{i,l}^{-1} (T^j-\tilde{\alpha}_{353,\;j} M^{353})_l,
\label{eq:chisq_alpha}
\end{equation}
and the indexes $i$ and $l$ are running from 1, $n_{groups}$ in the bin to which the pixel $\vnh$ belongs. The matrix $\vN^{-1}$ corresponds to the noise inverse covariance matrix, which is diagonal if $N$ is poissonian noise. The estimate and error associated to $\tilde{\alpha_{353,\;j}} $ are given by
\begin{eqnarray}
\tilde{\alpha}_{353,\;j} & = & \frac{ \sum_{i,l} T^j_i\vN_{i,l}^{-1} M^{353}_l }
{\sum_{i,l} M^{353}_i \vN_{i,l}^{-1}  M^{353}_l }, \\
\sigma^2_{\tilde{\alpha}_{353,\;j}} & = & \frac{1 }
{\sum_{i,l} M^{353}_i \vN_{i,l}^{-1}  M^{353}_l } .
\label{eq:al_sgal}
\end{eqnarray}

\begin{figure*}
\centering
\plotancho{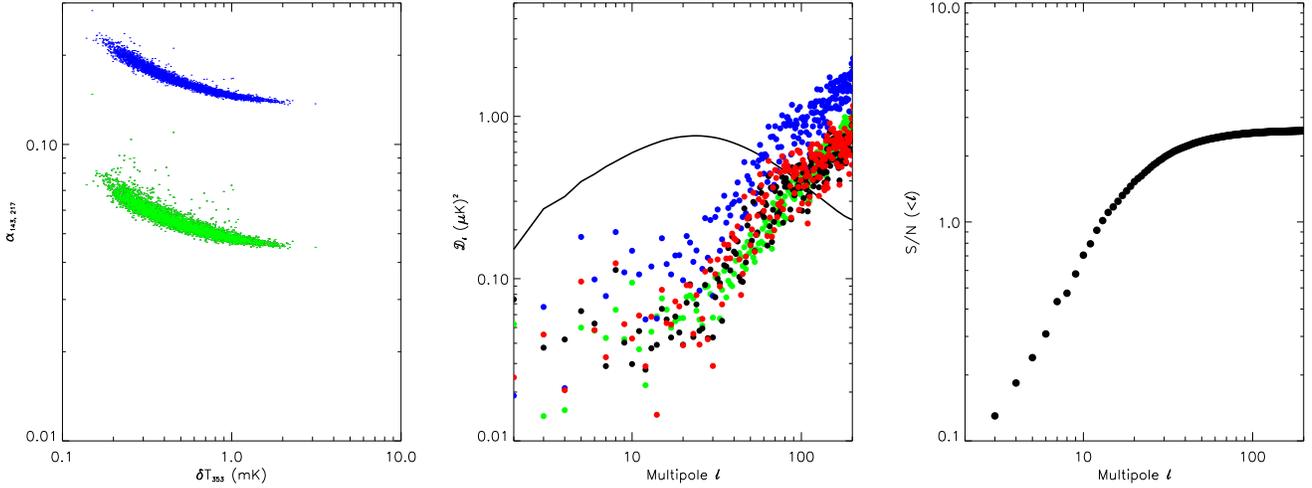}
\caption[fig:galaxy1]{{\it Left panel:} Effective spectral coefficient between 353 GHz, on the one hand, and 143 GHz (green points) and 217 GHz (blue points). {\it Middle panel:}  The expected ISW -- tSZ cross power spectrum amplitude (at 143 GHz, after masking all clusters below $z<0.3$) given by the solid black line is compared to the residuals present in the 143 GHz (green circles) and 217 GHz (blue circles) channels, respectively. The effective bias in the ISW -- tSZ cross spectra are displayed by red circles, while the black ones display the increase in the error budget (last term in Eq.\ref{eq:error_res}). Units are expressed in ${\cal D}_l \equiv l(l+1)C_l / (2\pi)$. {\it Right panel:}
  Cumulative S/N of the ISW -- galaxy cross power spectrum below a
  given multipole $l$.  }
\label{fig:galaxy1}
\end{figure*}

The residuals given by Eq.\ref{eq:model_res} will bias the tSZ -- ISW correlation estimates, and also increase their errors. The former is given by the second term in right hand side of this equation:
\begin{equation}
E\left[ \langle (a^j_{l,m}-a^{217}_{l,m}) (a_{l,m}^{217})^*\rangle \right] = C_l^{ISW-tSZ_j} + 
 \langle (a^{j,\; res}_{l,m}-a^{217, \;res}_{l,m} ) (a_{l,m}^{217,\; res})^* \rangle,
 \label{eq:bias_cross}
 \end{equation}
 where the multipoles of the residuals at channel $j$ (Eq. \ref{eq:model_res}) are given by the $a^{j,\; res}_{l,m}$-s. The impact of residuals on the dispersion of the tSZ -- ISW correlation estimates is modeled as in Eq.\ref{eq:ClISWSZvarianceIdeal}:
\begin{equation}
\Delta [C_l^{ISW-tSZ_j}] = (C_l^{ISW-tSZ_j})^2 + C_l^{CMB} (C_l^{tSZ_j,\;1h} + C_l^{tSZ_j,\;2h} + C_l^{j-217,\; res}),
\label{eq:error_res}
\end{equation}
with $C_l^{j-217,\; res}$ the power spectrum multipole of the residual difference map $\delta T^j_{res} - \delta T^{217}_{res}$ and where we have assumed that the power spectrum of residuals at 217 GHz is much smaller than the CMB power spectrum ($C_l^{217,\; res}\ll C_l^{CMB}$). As before, the S/N for a given multipole will be given by $(S/N)^2_l =  ( C_l^{ISW-tSZ_j} )^2 / \Delta [C_l^{ISW-tSZ_j}]$.

\begin{figure}
\includegraphics[width=8cm]{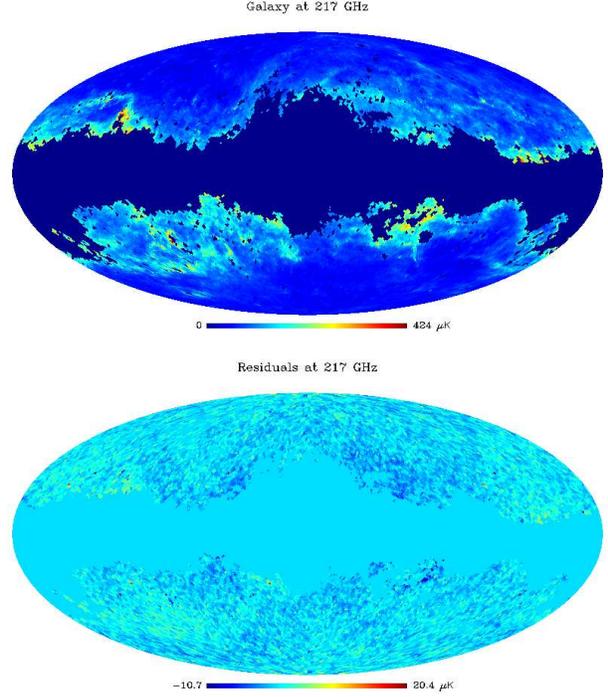}
\caption{{\it Top panel:} Total galactic emission outside the mask covering 40\% of the sky at 217 GHz in thermodynamic temperature units. {\it Bottom panel:} Residuals outside the mask at 217 GHz after using a perfect dust tracer at 353 GHz and assuming a white noise level of 80 $\mu$K at scales of 5 arcmins.}
\label{fig:galaxy2}
\end{figure}

We computed the total S/N for the ISW -- tSZ cross correlation for different levels of sky coverage $f_{sky}$, bin size $n_{group}$ and Gaussian white noise amplitude $\sigma_N \equiv \sqrt{ \langle N_{ii}\rangle }$. We considered the case where all clusters below $z<0.3$ are masked out, since it is the one giving rise to highest S/N in the ideal (contamination free) case, and a low frequency of 143 GHz. We first assumed a noise amplitude of $\sigma_N = 80 \;\mu K$ in square pixels of 5 arcmin size, and checked for the resulting S/N after binning in groups of varying lenght $n_{groups}$. This noise amplitude is above the predicted values for HFI channels 100, 143 and 217 GHz, for which values of 14.2, 9. and 14.6 $\mu$ K in pixels of size 5 arcmins are expected. One would guess, however, that any component separation algorithm (required to remove an estimation of the CMB in each channel) would sensibly increase the overall noise level.
We found that after binning pixels in groups of size $n_{groups} = 1000$ and masking 40\% of the sky, residuals would drop below the  ISW -- tSZ cross correlation amplitude for the 143 and 217 GHz channels. In the middle panel of Fig.\ref{fig:galaxy1} green circles display the amplitude of dust residuals at 143 GHz, while those at 217 GHz are given by blue circles. The effective bias on the ISW -- tSZ cross correlation (which is given by the solid thick line) are shown by red circles. The extra (last) term in the error computed in Eq.\ref{eq:error_res} is displayed by black circles. In all cases we are showing {\it pseudo-}power spectrum multipoles. In this configuration, the total S/N achieved was approximately 2.6 (as shown by the right panel of Fig.\ref{fig:galaxy1}). {When comparing to Fig.6  of \citet{leach}, we see that our level of residuals is comparable to those obtained after using the whole set of {\it Planck} frequencies for component separation}. The accuracy of our simple cleaning procedure is displayed by Fig.\ref{fig:galaxy2}: the top panel shows the total contaminant emission at 217 GHz, while the bottom one  displays the residuals after the cleaning procedure described above was applied.

We also explored how S/N depends on $\sigma_N$ and $n_{groups}$, see Fig. \ref{fig:galaxy3}. The top panel shows how S/N varies with $\sigma_N$ for different values of $n_{groups}$: black, red, green and blue lines correspond to $n_{groups}$=1,10, 100 and 1000, respectively. In the bottom panel, those same colors correspond to $\sigma_N = $ 1, 10, 50 and 100 $\mu$K respectively. The top panel displays how S/N is degraded as the noise level is increased, being this degradation more important for low values of $n_{groups}$ for the same value of $\sigma_N$. However, in the cases of low noise, the reconstruction of dust is more accurate for small $n_{groups}$ and higher values of S/N  are reached in such cases. The bottom panel, instead, shows how increasing the bin width ($n_{groups}$) improves the S/N only in the regime where noise is dominating: for low noise levels binning only degrades the quality of the subtraction. Note that in ideal conditions ($n_{groups}=1$, $\sigma_N=0$ $\mu$K), S/N$\simeq 3.9$ for $f_{sky}$=0.6 considered here.

\begin{figure}
\includegraphics[width=8cm]{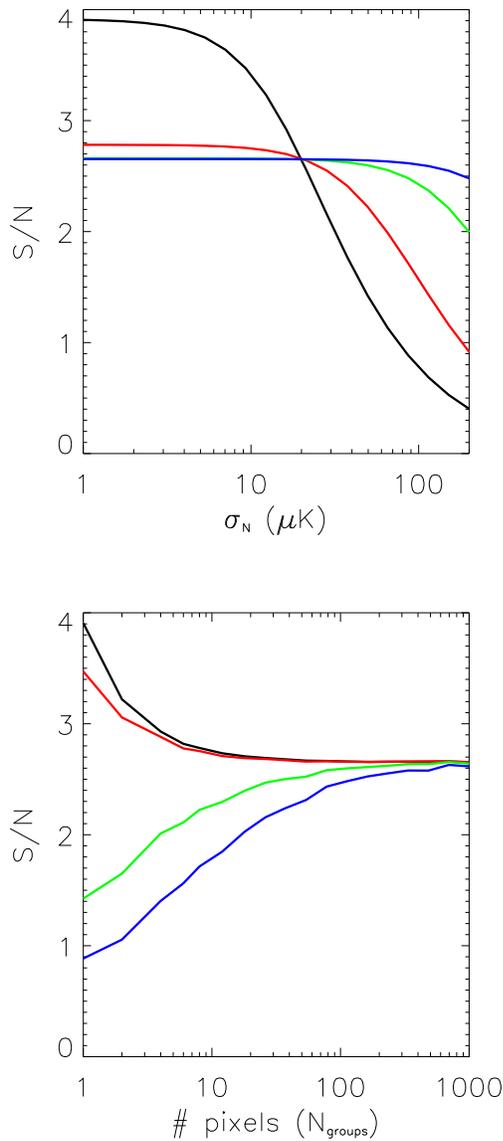}
\caption{{\it Top panel:} S/N versus residual effective noise for different values of $n_{groups}$:  black, red, green and blue lines correspond to $n_{groups}$=1,10, 100 and 1000, respectively. {\it Bottom panel:} S/N versus bin width $n_{groups}$: black, red, green and blue lines correspond to $\sigma_N = $ 1, 10, 50 and 100 $\mu$K respectively.}
\label{fig:galaxy3}
\end{figure}

\section{Discussion and conclusions}

The problem of component separation in multifrequency CMB observations for experiments like {\it Planck} has been a subject of active investigation \citep[e.g.,][]{stompor09,leach,eriksen08,juan07,vlad05}. The different contaminants (either galactic and extragalactic) show generally a different spectral dependence when compared to the CMB. On the low frequency side, foreground powerful in radio wavelenghts fall steeply and become the sub-dominant foregrounds at frequencies above $\nu \sim 70$ GHz, \citep{forewmap1}. Above this frequency, the presence of dust absorbing UV radiation and re-emitting it in the sub-millimeter and millimeter range constitutes the dominant contaminant. An experiment like {\it Planck}, with four high-angular resolution channels covering the 217 -- 857 GHz frequency range, should provide an accurate description of this foreground. With these data in hand, currently existing models describing the physics of dust emission should be improved further and accurate extrapolations to lower frequencies should be enabled. 

In our toy model describing the impact of the contaminants we assumed that the 353 GHz channel was a perfect (CMB -- free) dust tracer, and that the signal at low frequencies (100 or 143 GHz) was either due to dust or a white Gaussian noise signal. These assumptions may be overly optimistic (in regard to the 100 \& 217 GHz channels themselves), but however a generic combination of all nine channels (ranging from 30 GHz up to 857 GHz) are expected to provide an estimation of each of the components (CMB + foregrounds) whose residuals are expected to lie close to the level of $\sim 0.1$ ($\mu$K)$^2$ \citep{leach}. This is already the accuracy ballpark that our simple analysis proved to define the regime of detectability of the tSZ -- ISW cross correlation, and there may still be room for a more optimized channel combination oriented to unveil the particular tSZ--ISW cross correlation. The use of high galactic latitude HI maps as tracers of galactic cirrus could allow to lower the impact of those residuals at the level of a percent in $C_\ell$, (G. Lagache, private communication). \cite{NestorFernandez08} computed the cirrus power spectrum at different frequencies and different HI column densities. They found, in the case of fields that have a very low level of dust contamination, that the cirrus power spectrum at 217 GHz is of the order of 5 ($\mu$K$_{\rm RJ})^2$ at $\ell=10$ in units of $l(l+1)C_l/(2\pi)$. A one percent residual of the cirrus emission is thus smaller than 0.05 ($\mu$K$_{\rm RJ})^2$, i.e., 0.4 ($\mu$K$_{\rm CMB})^2$. Current foreground residual estimates based upon the works of \cite{NestorFernandez08} and \cite{leach} suggest that at the frequencies of
interest (100 -- 217 GHz) the contaminant residuals remain a factor of a few above our requirements. How much room there is for improvement below those limits is
something yet to be estimated from real data.

{\it Planck} data at high frequencies provide a more profound knowledge of the dust properties in both our galaxy and extragalactic sources, together with the mechanisms involving its emission in the sub-millimeter range. It is nevertheless important to bear in mind that, since the different frequencies sample the infrared galaxy populations {at different redshifts} and the galaxy linear bias evolves with redshift \citep[e.g.][]{Lagache07,VieroBlast09}, using high frequency maps so as to clean the 143 and 217 GHz could potentially degrade the residual level we obtained in Section 5 with our template fitting method. This issue is under current investigation within the {\it Planck} collaboration.

Even in the worst scenario in which foreground residuals are too high and complicated to prevent the detection of the tSZ -- ISW cross correlation, the upper limits to be imposed on it are of cosmological relevance, since it would constrain cosmological parameters like $\sigma_8$, $\Omega_m$ or $\Omega_{\Lambda}$.

We have shown that the tSZ -- ISW cross correlation constitutes a CMB contained test for Dark Energy. The peculiar frequency dependence of the tSZ effect and the availability of multi-frequency all sky CMB observations provided by the experiment {\it Planck} should enable an estimation of this cross correlation provided the hot gas is a fair tracer of the potential wells during the cosmological epochs where the ISW is active. Our theoretical study shows that the Poisson/shot noise introduced by the modest number of very massive, very bright in tSZ galaxy clusters can be attenuated by masking out those tSZ sources below redshift $z< 0.3$. In the absence of a massive galaxy cluster catalog below that redshift, it would suffice to excise from the analysis those tSZ clusters clearly detected by {\it Planck} in order to achieve S/N of the order of 3.9 ($f_{sky}=1$).  This tSZ -- ISW cross correlation detection would not require the use of deep-in-redshift and wide-in-angle galaxy surveys, but only the combination of different frequency CMB observations. This would hence provide a different approach for ISW detection with different systematics to other attempts based upon CMB -- galaxy survey cross correlations. If foreground residuals are kept at or below the $\sim 0.04 $ ($\mu$K)$^2$ level (in $l(l+1)C_l/(2\pi)$ units at $l\sim 10$) in the frequency range 100 -- 217 GHz, then the tSZ -- ISW correlation should provide a valid and independent test for the impact of Dark Energy on the growth of structure and the evolution of large angle CMB temperature anisotropies.

\section*{Acknowledgments}
NT is grateful for hospitality from the Max-Planck Institut f\"ur Astrophysik in Garching where part of this work was done. NT warmly thanks Guilaine Lagache, Aur\'elie Penin and Mathieu Langer for useful discussions that helped improving this paper. CHM is also grateful to the Institut d'Astrophysique Spatiale (IAS) for its warm hospitality during his frequent visits to Orsay.
\label{lastpage}

\bibliographystyle{mnras}
\bibliography{biblio}
\end{document}